# Wavefront Modulation and Subwavelength Diffractive Acoustics with an Acoustic Metasurface

Yangbo Xie[1], Wenqi Wang[1], Huanyang Chen[2], Adam Konneker[1], Bogdan-Ioan Popa[1]

and Steven A. Cummer[1*]

1. Department of Electrical and Computer Engineering, Duke University, Durham, North Carolina 27708, USA
2. Department of Physics, Soochow University, 1 Shizi Street, Suzhou 215006, People's Republic of China

**Abstract**

Metasurfaces are a family of novel wavefront shaping devices with planar profile and subwavelength thickness. Acoustic metasurfaces with ultralow profile yet extraordinary wave manipulating properties would be highly desirable for improving the performance of many acoustic wave-based applications. However, designing acoustic metasurfaces with similar functionality as their electromagnetic counterparts remains challenging with traditional metamaterial design approaches. Here we present a design and realization of an acoustic metasurface based on tapered labyrinthine metamaterials. The demonstrated metasurface can not only can steer an acoustic beam as expected from the generalized Snell's law, but also exhibits various unique properties including surface wave conversion, extraordinary beam-steering and apparent negative refraction through higher-order diffraction. Such designer acoustic metasurfaces provide a new design methodology for acoustic signal modulation devices and may be useful for applications such as beam-steering, surface wave manipulation, high efficiency sound absorption, acoustic imaging and ultrasound lens design.

Correspondence and requests for materials should be addressed to S.A.C. (email: cummer@ee.duke.edu)

Recent years have witnessed the emergence of a family of planar wave modulation devices, known as metasurfaces[1-6]. These engineered surfaces of subwavelength thicknesses are capable of many forms of wave manipulation and thus have drawn significant attention from both physics and engineering communities. Not only do these metasurfaces inspire the revisiting of fundamental physical laws governing wave propagation, but also they promise a bright future for useful devices with extremely compact footprint.

Electromagnetic metasurfaces have been experimentally realized with optical antennas or microwave metamaterials[1-6]. However, mapping the success of electromagnetic metasurfaces to the acoustic domain is challenging, mostly due to the mechanical nature and distinct properties of acoustic waves. Some initial theoretical exploration[7, 8] has recently been shown, nevertheless, only limited experimental investigations have been demonstrated[9].

In this work, we demonstrate an acoustic metasurface capable of general wavefront modulation through four distinct wave manipulation properties: anomalous refraction, surface wave conversion, extraordinary beam-steering and apparent negative refraction through higher order diffraction. We verify that the generalized Snell's law provides an accurate description for the first two properties. The last two properties, in contrast, are produced from the interaction of the periodicity of the metasurface and the spatially varying phase modulation, creating an additional term in the generalized Snell's law. Our metasurface is essentially a metamaterial/phononic crystal hybrid structure[10] and both the local and non-local responses can be leveraged for the purpose of shaping the wavefront with a much larger degree of control than traditional acoustic wave manipulation methods. Such thin planar acoustic metasurfaces, along with existing bulk metamaterials[11-20], provide a new design methodology for acoustic wave modulation, sensing, and imaging systems.

**Results**

**Engineering surface phase gradient and anomalous refraction**

To design a metasurface with subwavelength thickness, planar profile and nearly arbitrary wavefront modulation, demanding requirements need to be placed upon the material parameters. A recently proposed category of acoustic metamaterials, termed labyrinthine metamaterials[14, 21, 22], shows promise as building blocks for such wavefront-shaping metasurfaces. This family of unit cells feature high energy throughput, broad complex (amplitude and phase) modulation range and non-resonant dispersion. Under the guidance of the generalized Snell's law[1], we therefore designed our metasurface based on a series of tapered labyrinthine metamaterials[22]. The classical Snell's law rests upon the assumption that the accumulated phase is continuous across the interface. However, introduction of abrupt phase variation generalizes Snell's law with an extra term describing the phase gradient on the surface ($\xi = \frac{d\phi_s}{dx}$):

$$(sin\theta_t - sin\theta_i)k_0 = \xi \qquad (1)$$

By engineering the phase gradient term, nearly arbitrary wavefront modulation can be realized—including focusing, beam steering and surface wave conversion to name a few.

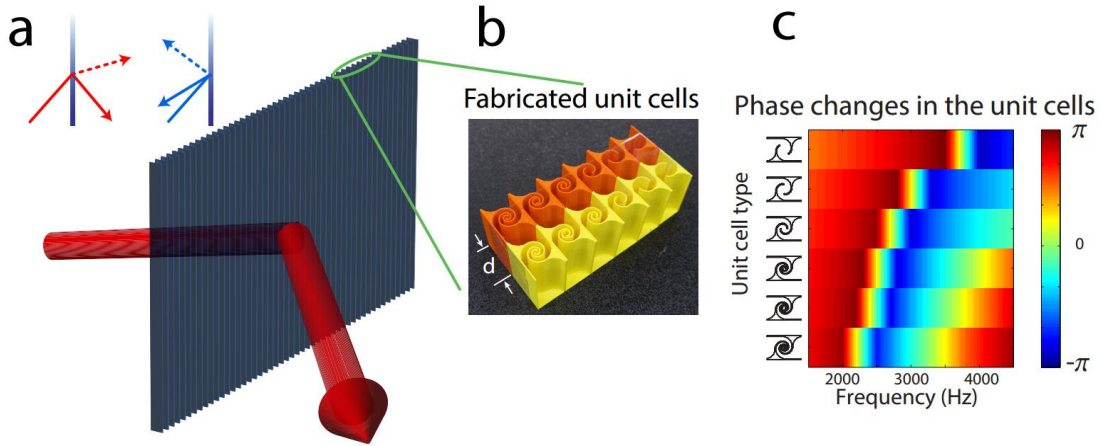

**Figure 1 | Wavefront modulating thin planar metasurface and its constituting elements.** (a) Concept schematic of a planar acoustic metasurface and the generalized Snell's law. Dashed arrow line indicates the positive refraction/reflection ray (corresponding to the modulation effect described by the generalized Snell's law) and solid arrow line indicates the negative refraction/reflection ray (corresponding to the hybrid response with the contribution from both the local and nonlocal effects). (b) Photograph of the fabricated unit cell prototypes. (c) 6 types of unit cells and the corresponding phase changes through the unit cells.

Figure 1b and Figure 1c showcase six types of tapered labyrinthine unit cells[24] with different complex modulation. This set of unit cells have uniform thickness (d in Figure 1b) that is around 25% to 35% of the wavelength of the interested frequency range (2500Hz to 3500Hz) and the difference of their modulated phase change can cover a complete $2\pi$ range with single layer (reflective case) or double layers (transmissive case). Figure 1c plots the phase change in each type of unit cell over the interested frequency range. At 3000Hz, the difference of phase change between adjacent types of unit cell is about $\pi/6$ and the phase difference is roughly preserved over a broad bandwidth. (This broadband effect will be explored in an upcoming publication). A transmissive metasurface composed of this set of unit cells approximates a phase gradient of $2\pi/\Gamma$, where $\Gamma$ is the spatial period of the metasurface. The generalized

Snell's law predicts the transmitted angle to be $\theta_t = \mathrm{asin}(sin\theta_i + \frac{\xi}{k_0})$, although as we show below, an additional term can be important due to the periodicity of the structure of the metasurface.

We first examine the case of normal incidence, i.e., $\theta_i = 0$. Here two metasurfaces with different gradient of phase $\xi$ were investigated. The measurement results for $\xi = 0$, $\xi = 3.3 (2\pi \cdot rad \cdot m^{-1})$, $\xi = 6.7 (2\pi \cdot rad \cdot m^{-1})$ were compared with the theoretical transmitted angles calculated with the generalized Snell's law (shown as the blue solid curve in Figure 2b) and excellent agreement is found. The physics of oblique incidence is much richer with surface mode playing a more important role; these effects will be explored in the subsequent parts of this paper.

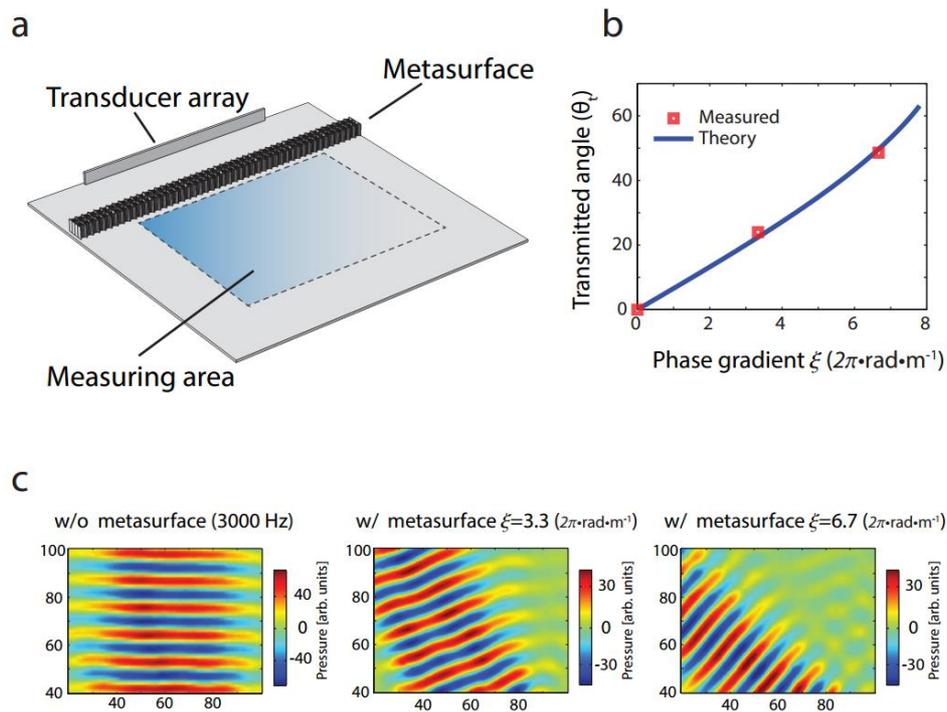

**Figure 2 | Anomalous refraction measurements**. (a) Experimental setup. The top plate of the waveguide (not shown in the diagram) is located 50.8mm above the bottom plate. (b) Relation between the transmitted angles and the phase gradients of the metasurfaces at normal incidence. Blue solid line is

the theoretical prediction from $\theta_t = \text{asin}(\frac{\xi}{k_0})$ and red squares are the measured values. The corresponding measured field patterns for these three phase gradients are shown in (c).

**Conversion from propagating to surface wave**

When the incident angle $\theta_i$ approaches a certain critical angle $\theta_c$, the transmitted wave will bend towards the surface and further increasing $\theta_i$ will generate surface wave whose surface wavenumber $k_x > k_0$ and the wavevector component perpendicular to the surface has an imaginary value $k_z = i\sqrt{k_x^2 - k_0^2}$. A theoretical estimation using a simple model with continuously varying phase modulation gives the value of the critical angle as $\theta_c = \text{asin}\left(1 - \frac{\xi}{k_0}\right) = 13.7°$ for the case of $\xi = 6.7$ $(2\pi \cdot rad \cdot m^{-1})$. As an example of the surface wave conversion, we pick $\theta_i = 25°$ which is located around the center of the surface mode range (blue region) of Figure 4a. In this case, the generalized Snell's law predicts the surface wavenumber to be $k_x = k_0 sin\theta_i + \xi \approx 10.4$ $(2\pi \cdot rad \cdot m^{-1})$, which is larger than the free-space wavenumber $k_0 \approx 8.75$ $(2\pi \cdot rad \cdot m^{-1})$. The measured surface wavenumber agrees with the prediction from the generalized Snell's law within about 7%. The measured field pattern for $\theta_i = 25°$ is shown in Figure 3a where the non-radiating surface wave can be clearly seen. Figure 3b shows the spatial Fourier transform of the measured near field along the surface and show that the modulated wave has a transverse wave vector larger than that of the free-space propagation. Such property of converting propagating wave to surface wave can be useful for efficiently coupling acoustic waves to surface modes and absorbing acoustic energy.

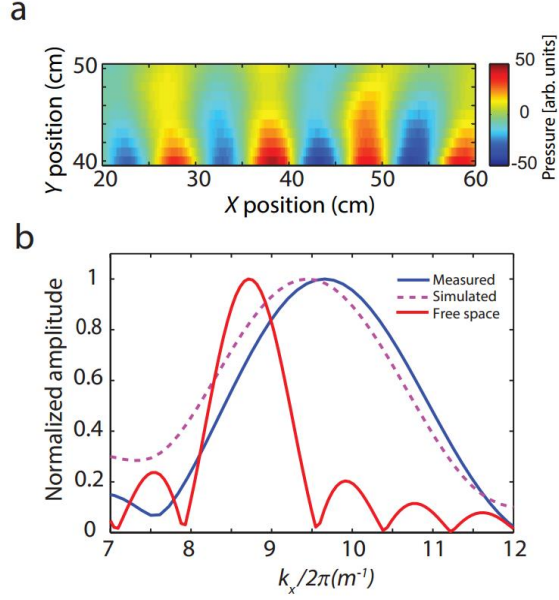

**Figure 3 | Measurements of surface wave converted from propagating wave.** (a) Measured near-field surface wave pattern for $\theta_i = 25°$. (b) Spatial Fourier transform of the measured near-field pressure along the metasurface, compared with the simulated result and reference free-space wavevector to prove the evanescent nature of the converted wave.

**Extraordinary beam steering and apparent negative refraction**

When the period of the structure is comparable with wavelength, non-local effects from periodicity will play a crucial role in wave propagation, and the generalized Snell's law of refraction/reflection will become wave-vector dependent. Under this condition, equation (1) can be reformulated as

$$(sin\theta_t - sin\theta_i)k_0 = \xi + n_G G \quad (2)$$

where $G$ is the amplitude of the reciprocal lattice vector. The structure can be regarded as a phononic crystal whose lattice element is a set of inhomogeneous and dispersive metamaterial unit cells[10]. In one-dimensional case, isofrequency surface analysis can be used for predicting the diffractive properties of the incident wave.

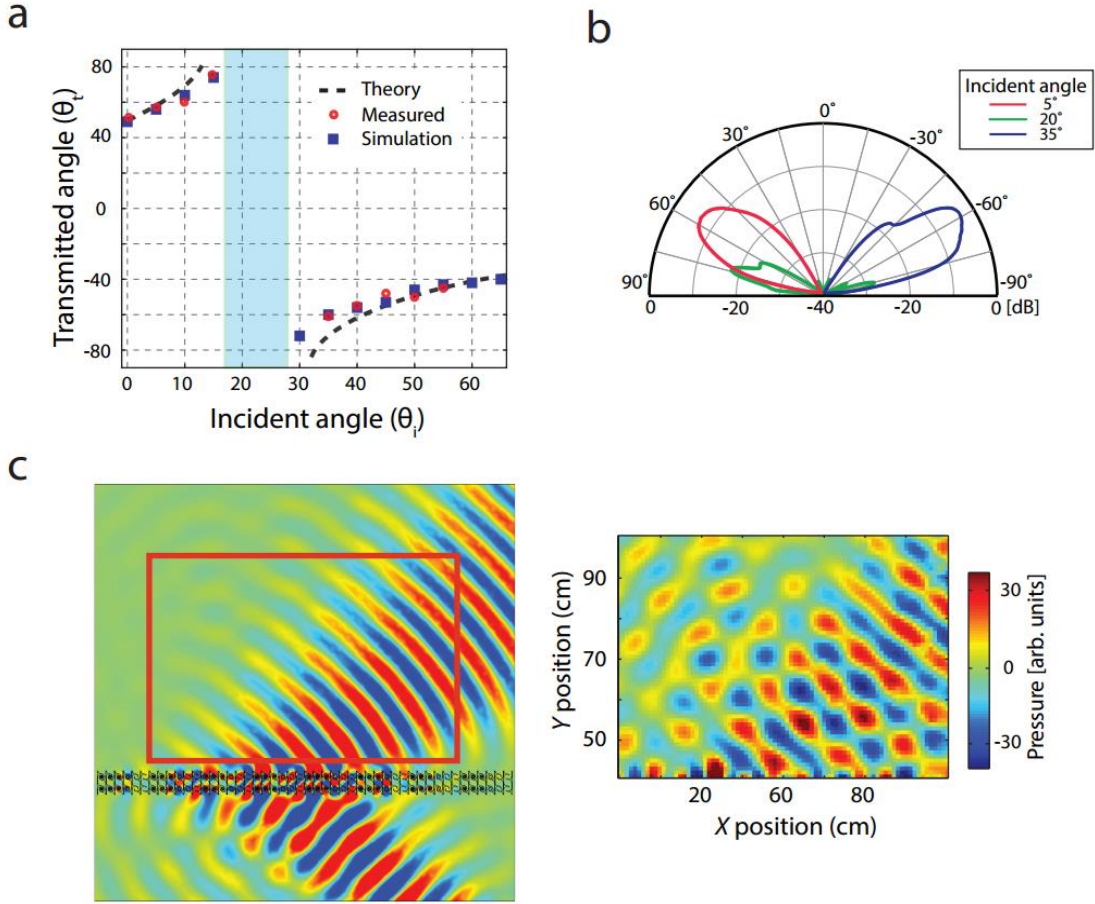

**Figure 4 | Extraordinary beam-steering and negative refraction.** (a) Relation between incidence angle θ$_i$ and transmitted angle θ$_t$. The theoretical curve is calculated based on the generalized Snell's law with the non-local term $n_G G$ taken into consideration (equation (2)). (b) Polar plot of the simulated radiation patterns for incident angle of 5° (smaller than θ$_c$), 35° (larger than θ$_c$) and 20° (around θ$_c$). (c) Field pattern at 45° (left: simulation, right: measurement), demonstrating the negatively refracted transmission when θ$_i$ > θ$_c$.

Figure 4a plots the measured $\theta_t \sim \theta_i$ data and compares it with the simulated one as well as the theoretical predication. The theoretical curve is calculated from equation (2) where a non-local term $n_G G$ is integrated into the generalized Snell's law. For an incident angle smaller than the critical angle $\theta_c$, it can be proven with classical diffraction theory that $n_G = 0$. However, when $\theta_i > \theta_c$ (the right side of the blue region in Figure 4a), $n_G$ will jump to a non-zero value ($n_G = -3$ for our design) as required by the surface momentum matching condition. Therefore, the $\theta_t \sim \theta_i$ curve exhibits an abrupt change when $\theta_i$ crosses the non-radiating region and a very small perturbation of the input signal $\theta_i$ will lead to a significant shift in the output signal $\theta_t$. This phenomenon is related to the passing off orders found in diffractive grating anomalies[23, 24]. Potential applications of such extraordinary beam-steering effect include beam forming, acoustic switching and surface wave sensing[25]. It is also worth noting that in the non-local regime (the right side of the blue region in Figure 4a), the transmitted beam exhibits negative refraction (refracted beam is on the same side of the incident beam with respect to the surface normal). This is different from the traditional negative refraction phenomena which are usually obtained with bulk negative refractive index metamaterials or phononic crystals, our metasurface achieves the same effect with the modulation from the subwavelength thin interface.

**Discussion**

In conclusion, we have designed and demonstrated an acoustic metasurface which produces arbitrary complex modulations of an incident wavefront. Based on tapered labyrinthine unit cells, the metasurface is shown to efficiently redirect incident acoustic beams as described by the generalized Snell's law. We also demonstrate that, when illuminated beyond the critical angle, the metasurface can also convert an incident propagating wave to a non-radiating surface wave. Last, we demonstrate a regime where periodicity of the metasurface also influences the transmitted energy, creating effects such as extraordinary beam-steering and apparent negative refraction through higher order diffraction. Our work opens the door to the arena of subwavelength diffractive acoustic devices which can be designed to achieve nearly arbitrary complex modulation of the wavefront. With these ultralow profile planar acoustic metasurfaces, we can expect to see novel acoustic signal modulation, sensing and imaging applications in the near future.

**Method**

**Field mapping measurement**

The unit cells were made of acrylonitrile butadiene styrene (ABS) thermal plastics and were fabricated using fused filament fabrication (FFF) 3D printing technology. The field mapping measurement was performed in our lab-made two-dimensional acoustic waveguide[26, 27]. The experimental setup is depicted in Figure 2a. A linear transducer array was used to generate a plane wave with a Gaussian amplitude profile along the wave front and a microphone was swept over the measuring area by a two-dimensional linear stage to record the transmitted wavefronts.

**Numerical simulations**

The simulation was performed with the commercial finite element analysis solver COMSOL Multiphysics. The unit cell design was based on a standard effective acoustic parameter retrieval method[28]. The pressure field pattern simulation was performed by approximating the experimental conditions of the two-dimensional waveguide measurement system. The boundary conditions of the calculating area were set to the plane wave radiation boundary.

**Acknowledgments**

This work was supported by a Multidisciplinary University Research Initiative from the Office of Naval Research (grant N00014-13-1-0631).

**Author contributions**

Y.X. designed and performed the theoretical and numerical calculation. Y.X. and W.W. conducted the experiments. H.C. helped with the theoretical analysis. A.K. and B.-I.P. assisted the experiments. Y.X. and S.A.C. prepared the manuscript. S.A.C. supervised the project.

**Competing financial interests**

The authors declare no competing financial interests.